# Image restoration quality assessment based on regional differential information entropy

Zhiyu Wang[1,‡], Jiayan Zhuang [2,‡], Ningyuan Xu [3,*], Sichao Ye [2,*], Jiangjian Xiao [2,] and Chengbin Peng[1]

| 1 | Ningbo University, Ningbo 315211, China |
| 2 | Ningbo Institute of Industrial Technology, Chinese Academy of Sciences, Ningbo 315201, China |
| 3 | University of Chinese Academy of Sciences, Beijing 100049, China |
| * | Correspondence: yesichao@nimte.ac.cn |
| ‡ | These authors contributed equally to this work. |

**Abstract:** With the development of image recovery models, especially those based on adversarial and perceptual losses, the detailed texture portions of images are being recovered more naturally. However, these restored images are similar but not identical in detail texture to their reference images. With traditional image quality assessment methods, results with better subjective perceived quality often score lower in objective scoring. Assessment methods suffer from subjective and objective inconsistencies. This paper proposes a regional differential information entropy (RDIE) method for image quality assessment to address this problem. This approach allows better assessment of similar but not identical textural details and achieves good agreement with perceived quality. Neural networks are used to reshape the process of calculating information entropy, improving the speed and efficiency of the operation. Experiments conducted with this study's image quality assessment dataset and the PIPAL dataset show that the proposed RDIE method yields a high degree of agreement with people's average opinion scores compared to other image quality assessment metrics, proving that RDIE can better quantify the perceived quality of images.

## 1. Introduction

Image restoration is a long-standing and active area of research in digital image processing, including image denoising, deblurring, and super-resolution. Image restoration plays an important role in image understanding, representation, and processing. The goal of image restoration is to recover a clean potential image from a degraded image. However, while image restoration technology is achieving increasingly high-quality results, objective image quality assessment (IQA) metrics for restored images are not well aligned with subjective assessment metrics, which limits the development of the technology. Therefore, the design of objective IQA metrics to maintain consistency between subjective and objective assessments of restored images has become an important issue in image restoration.

Image recovery is a pathological inverse problem, and the infinite number of possibilities between the degraded image and the corresponding reference image determines the uncertainty of the problem. Traditional mean squared error loss methods for image restoration tend to generate the average of multiple potentially clean images. These methods yield high scores for mainstream objective assessment metrics such as the peak signal-to-noise ratio (PSNR), the mean squared error (MSE) between the original and degraded images, and the structural similarity index metric (SSIM) proposed by Wang et al. [1]. However, these methods are biased toward generating blurred and over-smoothed results, which leads to low perceived quality of the recovery results. To obtain clearer and more natural results, Johnson [2] proposed perceptual loss, which adds optimization of the model in the feature space, in contrast to the traditional optimization of the model in the original color space of the image.

This approach yields more similar image results at both the input and output levels and the feature levels. Ledig [3], Ramakrishnan [4], and Chen [5] also respectively proposed the use of generative adversarial networks to solve problems of image super-resolution, image deblurring, and image denoising and make the recovered images more consistent with the distribution of real images. Wang [6] combined both perceptual loss and adversarial loss, generating better results.Similarly. Ma et al. [7] used a gradient map as an additional guide to generate more realistic detailed textures. Although these methods yield high visual perceptual quality, they score low with respect to objective assessment metrics such as PSNR, MSE, and SSIM.

However, such objective assessment metrics are designed to compare the degree of pixel difference between the recovered image and the original image or the level of similarity between the two and do not correspond well to the perceived quality of the image. Over the past few years, an increasing number of scholars have adopted subjective assessment methods to properly evaluate the perceptual quality of restored images, using mean opinion scores (MOS) and relative mean opinion scores (DMOS) as the metrics for assessing restored images. The former judges the quality of an image by normalizing the observer's score, while the latter judges the quality of an image by normalizing the difference between the distortion-free and distorted images by the observer. However, it is time-consuming and sometimes impractical to obtain large-scale and valid subjective assessment results. Therefore, there is an urgent need for an effective IQA method.

Image information entropy [8], an IQA method originally proposed by Shannon to describe the uncertainty of the source, reflects the richness of image information from an information theory point of view and can better evaluate the perceptual quality of an image. Image information entropy can represent the amount of information contained in the aggregated features of the grayscale distribution of a grayscale image according to the following mathematical expression:

$$H = \sum_{i=0}^{255} p_i \log p_i \tag{1}$$

where $p_i$ is the proportion of pixels in the image with gray values to the total number of pixels. In general, the higher the information entropy is, the richer the content of the image is. However, traditional global information entropy does not provide a high degree of access to the structure and content of the image. Furthermore, the calculations required are time-consuming and require substantial computational resources.

Building on traditional image information entropy, this paper proposes a combination of the concept of regional information entropy and a reshaping of the calculation process for image information entropy using neural networks to overcome the shortcomings of traditional image information entropy and describe the degree of detail of the recovered image more intuitively. This approach can better quantify the perceptual quality of an image than metrics such as PSNR, SSIM, MSE, and MS-SSIM [9]. The contributions of this study are as follows:

1. In comparison to traditional IQA methods, the regional differential information entropy (RDIE) method proposed in this paper yields objective assessment results that agree better with subjective assessments.

2. Image information entropy is viewed and described from a new perspective, that is, neural network, which shows the possibility of simulating traditional algorithms by using convolution with specific weights and particular activation functions.

3. The traditional information entropy calculation method is serial, whereas the RDIE method proposed in this paper has a high degree of parallelism,Great improvement in computing speed.

## 2. Materials and Methods

*2.1. Related Works*

An ideal IQA method should be ast and reliable. IQA methods can be classified as subjective or objective depending on whether or not there is human involvement. Subjective quality assessment evaluates the quality of an image based on people's subjective perceptions, and since people may have different assessments of the same image, it is common practice to take the average of multiple people's assessments of distorted images as the assessment result. Objective IQA requires a mathematical model to calculate quantitative assessment results. Excellent IQA requires consistency between objective and subjective quality assessment scores. Objective quality assessment can be classified as full-reference image quality assessment (FR-IQA) or no-reference image quality assessment (NR-IQA), depending on the presence or absence of a reference image. The approach described in this paper is a full-reference objective IQA method.

2.1.1. No-reference method

The NR-IQA method involves quantifying the perceived quality of an image without a reference image, using only the image's own information for quality assessment. This method can be used in a wide range of scenarios because it is not limited by the reference image. Early NR-IQA methods were geared more toward specific types of distortion tasks. Ye et al. [10] obtained a feature dictionary by unsupervised feature learning, leading to the CORNIA method. Liu et al. [11] extracted natural statistical properties of distorted images in terms of structure, naturalness, and perceptibility, combined with unsupervised learning for IQA. Wang et al. [12] proposed a perceptual quality metric based on the Kullback–Leibler (KL) divergence of wavelet coefficient distributions for real images and scenes. The idea was further extended in subsequent studies [13,14,15,16] to quantify perceptual quality by various measures of deviation from natural image statistics in the spatial, wavelet, and neural-net-based deep features domains.

2.1.2. Full-reference method

FR-IQA uses the full image information to quantify image quality by assessing the degree of similarity between the image and the corresponding reference image. The early and most representative methods were MSE and PSNR, which calculate the difference between image and reference image pixels, but this approach does not take into account the human visual system's ability to perceive distortion differences, and thus inconsistent subjective and objective assessment results occur.

Compared to PSNR and MSE, the SSIM metric better reflects the quality of the restored image. SSIM assumes that human visual perception is adaptive in extracting structural information from the scene, so the luminance, contrast, and structural information between the distorted image and the reference image are measured separately, and the similarity is calculated, with higher scores being better. On this basis, Chen et al. [17] combined the gradient information of the image and proposed the gradient-based structural similarity metric. Wang proposed the MS-SSIM [9] based on multi-scale structural similarity comparisons. However, over the past few years, the requirement for more realistic detail in image restoration methods has increased, especially with the popularity of GAN-based image restoration methods, and there are still inconsistencies in the subjective and objective perceived quality of the SSIM and its associated metrics when evaluating the restored images.

Sheikh et al. [18,19] proposed the information fidelity criterion and visual information fidelity as metrics. These two methods have better consistency with the visual perception quality but have no response to the structure information of the image. In addition, the problem of sub-pixel mismatch between the restored image and the reference image is a key issue affecting the assessment of image quality. Kim et al. [20] propose eliminating sub-pixel level differences between images before assessing image quality. In addition, Liu et al. [21] found that the human eye is more sensitive to pixel points with high relative positional coherence and used phase matching for IQA. Zhang et al. [22] selected phase consistency and gradient information of interest to the human eye as features to assess image quality and compared the similarity between image features to assess image quality. In some cases, this yielded good agreement in subjective and objective image quality evaluation, but it still had difficulty in achieving the desired results in image restoration-type problems. By simulating the visual perceptual properties of the local perceptual field of HVS, Wu [23] divided image content into five regions: smooth regions, primary edges, secondary edges, regular textures, and irregular textures, and proposed a structure–texture decomposition approach based on perceptual sensitivity. This approach inspired this study because people perceive smooth, edge, and textured areas of an image very differently.

Recently, IQA methods based on deep neural networks have become popular, and these metrics are used as loss functions in image restoration problems, resulting in better image restoration results [24,25,26]. Despite the advances in IQA methods, only a few IQA methods (e.g., PSNR, SSIM, and PI) are regularly used to assess image recovery results.

*2.2. Method*

Traditional global information entropy calculates the ratio of each pixel to all pixels in an image to represent the amount of information contained in the aggregated features of the grayscale distribution in the image, which does not capture the structure and content of the image. Figure 1 (a) shows three images: a statue on the left, a rectangle in the middle, and stripes on the right. They look completely different, but according to the traditional global information entropy metric, they are identical, as they have exactly the same ratio of black to white pixels, 68:32. Compared to the global information entropy, the regional information entropy divides the image into several regions and calculates the information entropy separately, so that each region has a different structural content and can express the structural content of the image more clearly. So this paper uses regional information entropy to solve this problem, as shown in Figure 1 (b), after the regional information entropy processing, the three different pictures have obvious visual differences, and the regional information entropy captures the structure and content of the images very well.

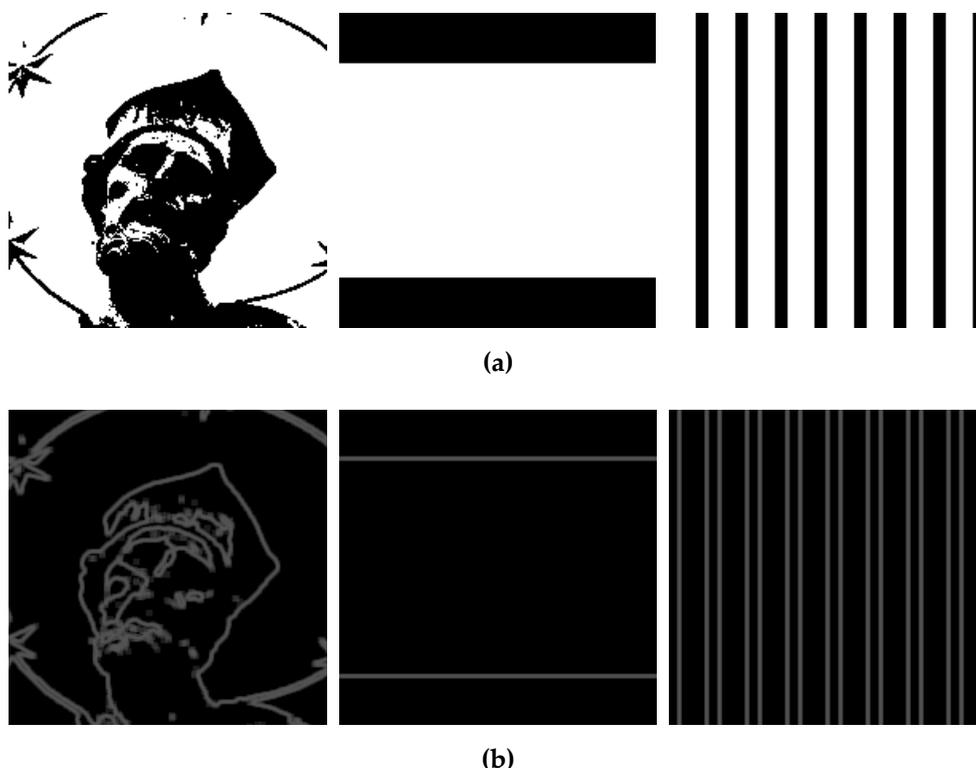

**Figure 1.** Comparison of regional information entropy before and after processing: (a) The statue on the left, the rectangle in the middle, and the stripes on the right have the same black and white proportions; (b) Results after regional information entropy processing.

Our method can be viewed as a series of transformations, transforming the image into a unique feature space and calculating their root mean squared error as their perception of similarity. The IQA method proposed in this paper is shown in Figure 2. The image restoration method is applied to the degraded image to generate the restored image. The regional information entropy method is applied to the corresponding reference image to obtain the corresponding regional information entropy feature map. The MSE between the feature maps is calculated as a quantitative result. The smaller the MSE is, the closer the recovered image is to the corresponding reference image in terms of information richness.

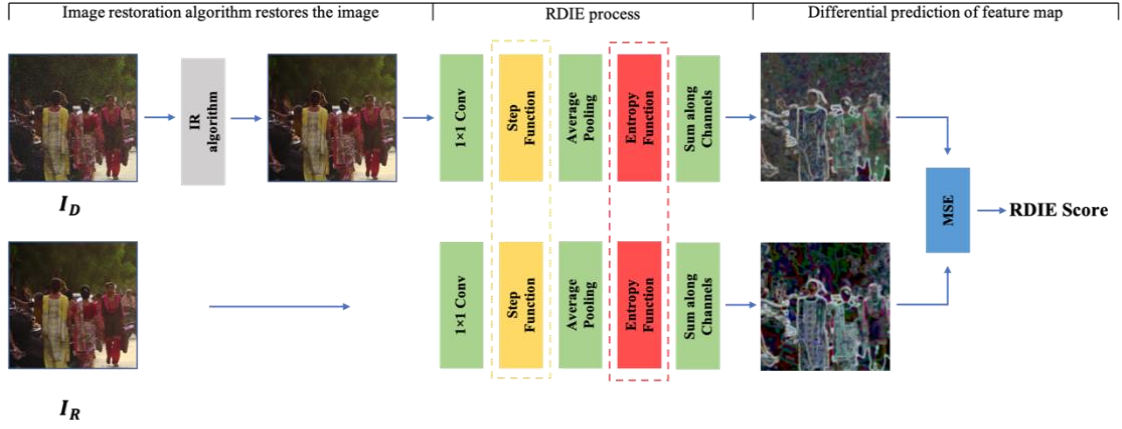

**Figure 2.** RDIE pipeline.

RDIE can be calculated as follows:

$$E(I, I^r) = L_2\big(M(I) - M(I^r)\big) \qquad (2)$$

$$= \sqrt{\frac{\sum_{y=0}^{\frac{height}{st}+st-h} \sum_{x=0}^{\frac{width}{st}+st-w} \big(H(R_{xy}) - H(R_{xy}^r)\big)^2}{\left(\frac{height}{st}+st-h\right)\left(\frac{width}{st}+st-w\right)}} \qquad (3)$$

where I is the test image, $I^r$ is the reference image, M(I) is the region information map of the test image, st is the stride of sliding windows, h is the height of the region, w is the width of the region, $R_{xy}$ is the region with upper left index (x, y) in the test image, and $H(R_{xy})$ is the information entropy of region $R_{xy}$. H(R) is defined as follows:

$$H(R) = -\sum_{n=0}^{L-1} P_l(R) \log_2 P_l(R) \qquad (3)$$

where L is the quantization level and $P_l(R)$ is the probability at a specific gray level l in the region, which can be defined as follows:

$$P_l(R) = \frac{1}{h*w} \sum_{i=0}^{h-1} \sum_{j=0}^{w-1} f_l(x_{ij}) \qquad (4)$$

where $\mathbf{x_{ij}}$ is the pixel value at (**x, y**) of the region and $\mathbf{f_l}$ is a piecewise function that can be defined as follows:

$$\boldsymbol{f_l(u)} = \begin{cases} \mathbf{1,} & \boldsymbol{u=1} \\ \mathbf{0,} & \boldsymbol{u \neq 1} \end{cases} \boldsymbol{u, l \in 0, 1, 2, \dots, L-1} \qquad (5)$$

Since the traditional use of sliding windows to calculate regional information entropy is very time-consuming, this paper uses neural networks to optimize the RDIE so that multiple windows can be used to process images in parallel to calculate image information entropy. As shown in the RDIE calculation process in Figure 2, this paper uses different channels to count the frequency of different grayscales, which allows each gray level to be calculated independently. In addition, this paper uses 1×1 convolutional layers, average pooling layers, and specific activation functions to form a neural network instead of the traditional method. The activation functions are the step function and the entropy function, shown below:

$$Step_L(x) = \begin{cases} 1, & 0 \leq x \leq \frac{256}{L} \\ 0, & else \end{cases} \quad (6)$$

$$Entropy(x) = \begin{cases} -x * log_2 x, & x \neq 0 \\ 0, & 0 \end{cases} \quad (7)$$

With the existing parallel computing platform, the computational efficiency of this paper is greatly improved compared to the traditional method. A comparison between the traditional method and the method proposed in this paper is shown in Table 1. The test selection was a 2040×1356×3 size image. The results indicate that $GIE_{nn}$ is approximately three times faster than $GIE_t$, while $RIE_{nn}$ is 5,400 times faster than $RIE_t$.

The main factors affecting the RIE are the quantization level $L$, the window dimensions $h$ and $w$, and the strides $st$. The quantization level affects the perception result mainly through the sensitivity to differences in pixel size. As shown in Figure 3, the two images have the same shape, the difference in brightness on the left is 1, while the difference in brightness on the right is 255, and the inconsistency between the two images is obvious to humans, with the right image having a much clearer boundary. The quantization using the traditional information entropy of 256 gray levels produces the same regional information entropy results for both images, which is clearly not in line with the perceived results. Therefore, this paper adopts a smaller quantization level.

Different window sizes affect the degree of detail in the image structure, and as shown in Figure 4, the results become clearer as the window size gets smaller and smaller.

Increasing the strides will reduce the impact of the image due to movement but will also increase the computational effort of the method.

Detailed experiments were conducted to explore the effects of different quantization levels, window sizes, and strides on RDIE, as described later in this paper.

**Table 1.** Speeds of traditional method and neural network method. RIE has a window size of 4×4 and a quantization level of 8. The subscript t denotes the traditional method, and nn denotes the method used in this paper.

| Metrics | $GIE_t$ | $GIE_{nn}$ | $RIE_t$ | $RIE_{nn}$ |
|---|---|---|---|---|
| Time/ms | 56.2 | 14.5 | 138600 | 25.9 |

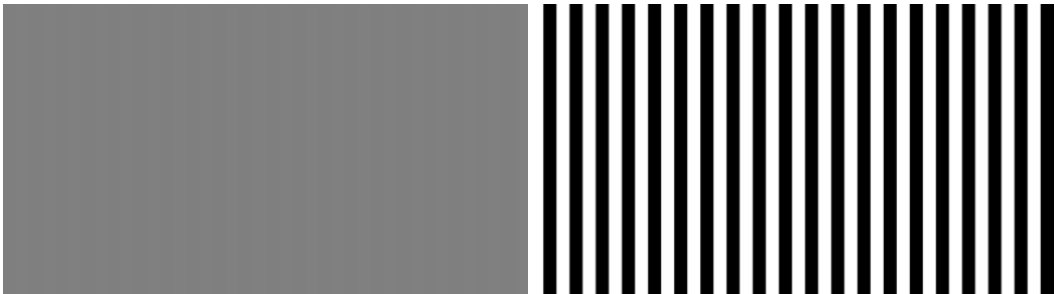

**Figure 3.** Stripe patterns in different colors. The difference between two adjacent stripes on the left is 1, while the difference on the right is 255.

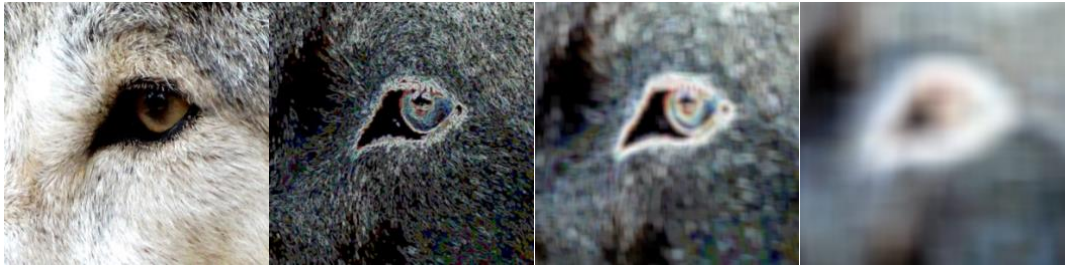

**Figure 4**. RIE results for different window sizes, from left to right: original image, generated by window size 4×4, generated by window size 16×16, generated by window size 64×64.

*2.3. Datasets*

2.3.1. Our Datasets

We studied the image super-resolution and denoising sub-problems of the image restoration problem using different data to produce IQA datasets. For the super-resolution sub-problem, we used images from the DIV2k [27] dataset. Because the images in the DIV2k dataset are too large and too numerous to be placed on an image quality manual evaluation page, 15 DIV2k images were selected and cropped to a 500×500 size to serve as reference images. For the image denoising sub-problem, images from the CBSD68 color dataset were used, with a Gaussian noise level of 50, the same as those used for the super-resolution sub-problem. Fifteen images were selected as reference images.

In this paper, Bicubic interpolation [28], EDSR [29], WDSR [30], SAN [31], SRGAN [6], and SPSR [7] are used as methods for SR to get high-resolution images from low-resolution images, DNCNN [32], FFDNet [33], IRCNN [34], IPT [35], and LIGN [36] as denoising methods to recover clean images from noisy images. In SR methods, Bicubic is a traditional up-sampling method. EDSR optimizes the SRResnet [3] network structure by removing the BN layer, reducing both computation time and optimizing recovery results. WDSR is based on EDSR, which expands the number of feature maps before the activation function in the Block, allowing the network to convey information better. SAN proposes a two-stage attention network, leading to stronger feature representation and feature relationship learning. SRGAN uses both perceptual loss and adversarial loss, making the perceptual quality of its restored images better than the previously mentioned methods. SPSR introduces gradient loss, which enhances the detailed texture of the restored image. Among the denoising methods, DNCNN is the first deep learning-based image denoising algorithm. FFDNet proposes a noise level map as input and noise estimation and noise images together as input to improve the generalization of noise. IRCNN trains a fast and efficient CNN denoising network and integrates it into a model-based optimization approach. IPT is based on the transformer's network model and differs from other image restoration methods in that it takes different head and tail sections for different restoration tasks, improving its accuracy for the corresponding image restoration task. LIGN proposes a layered input method that adds image gradient depth information to the network, enhancing edge and detailed texture regions, so it is more in line with people's perceived quality compared to the previous methods.

2.3.2. PIPAL

To evaluate the adaptability of RDIE to a wider range of image restoration tasks, this paper tests the method in this paper on the PIPAL [37] dataset. PIPAL is a huge IQA dataset containing 250 reference images, four subclasses, 40 distortion types, 29 thousand images, and 1.13 million human assessment scores. The four subclasses are traditional distortion, image super-resolution, denoising, and blending restoration. According to the research content of this paper, two subclasses of image super-resolution and denoising were selected as the research objects. The subclasses contain several existing model results of algorithms, which were divided into three categories for experimental study: traditional methods, PSNR-driven image restoration methods, and GAN-based image restoration methods.

PSNR-driven image restoration class algorithms are typically based on deep learning and produce outputs with sharper edges and better PSNR than traditional methods. The results of GAN-based image restoration class methods are more complex and challenging for IQA. They often contain similar but not identical texture details to the reference image and are difficult to assess effectively with IQA methods similar to PSNR.

This paper assesses the strengths and weaknesses of the IQA method by calculating the Spearman rank correlation coefficient (SRCC) between the IQA method and the subjective scores. This metric provides a good assessment of the monotonic correlation between IQA methods and people's perceived image quality, with the larger the absolute value of SRCC, the stronger the correlation.

**3. Results**

*3.1. Results of ablation experiments*

The main factors affecting RDIE are the sliding window size, quantization level, and stride. In this study, extensive experiments were done with PIPAL to find the optimal parameters, using a grid search method with window sizes from 2 to 16 and quantization levels from 2 to 80. In this paper, we define $RDIE_{s,L}$, where $s$ denotes the window size and $L$ denotes the quantization level. For example, $RDIE_{10,16}$ denotes an RDIE method with a window size of 10×10 and a quantization level of 16. Since it is difficult to display so many results directly, we only selected some of them for the purpose of drawing curves.

3.1.1. Different window sizes

As shown on the left side of Figure 5, the curves of the conventional image restoration method and the PSNR-driven image restoration method are very similar. They both reach their maximum value when the window size is 4. The GAN-based image restoration method has a relatively smooth curve and works best when the window size is between 5 and 8. Considering the balance of the three data types, 5 was selected as the optimal size for the sliding window.

3.1.2. Different quantization levels

As shown on the right side of Figure 5, as the quantization level increases, the SRCC starts with an upward trend, and the curve oscillates when it reaches about 20. Considering that the larger the quantization level is, the greater the consumption of computational resources is, so 32 was chosen as the best quantization level.

3.1.3. Different quantization levels

In theory, increasing the strides can reduce the effect of pixel misalignment on the results. In this paper, quantitative experiments were conducted on step sizes, as shown in Table 2, and different strides do have a small effect on SRCC, but the amount of computation increases geometrically as the step size decreases. Ultimately the paper chooses a step size equal to the window size in subsequent experiments.

**Table 2**. SRCC results for different strides.

| Stride | Traditional method | PSNR-oriented method | GAN-based method |
|---|---|---|---|
| 1 | 0.6484 | 0.7270 | 0.5284 |
| 2 | 0.6480 | 0.7205 | 0.5292 |
| 3 | 0.6474 | 0.7203 | 0.5282 |
| 4 | 0.6533 | 0.7255 | 0.5276 |
| 5 | 0.6476 | 0.7203 | 0.52279 |

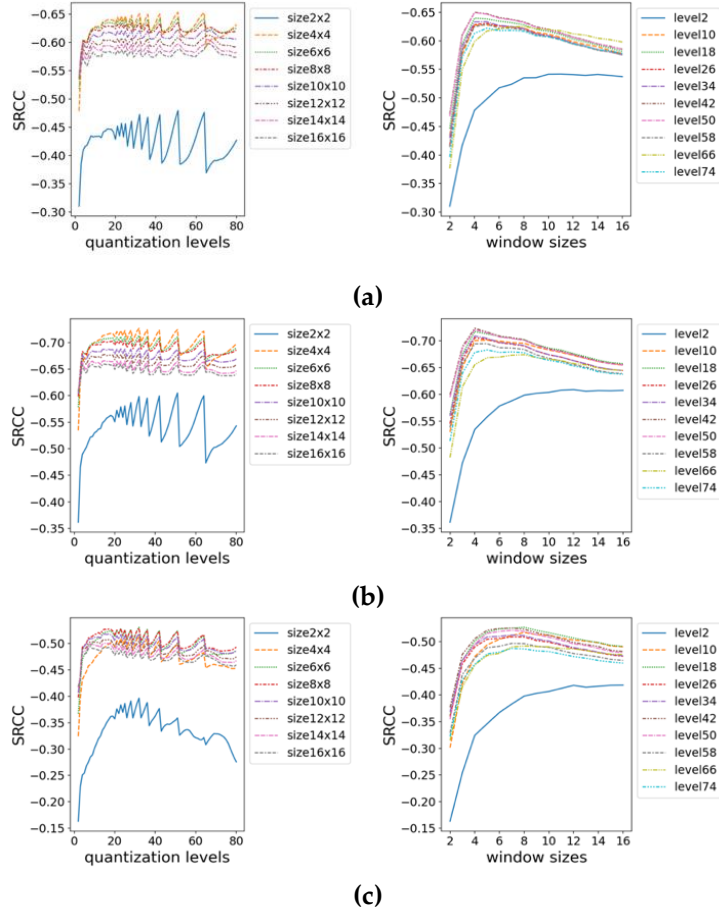

**Figure 5**. SRCC with different quantization levels and window sizes: (a) Traditional method; (b) PSNR-oriented method; (c) GAN-based method.

*3.2. Results in Our Dataset*

The results of the different restoration models are presented to the user, who rates the images from one to five stars based on perceived quality. Manual assessment results were collected to quantitatively evaluate the perceptual quality of images generated by different methods. The IQA dataset produced in this study contains a total of 30 images; 11 image recovery methods, i.e., six SR methods and five denoising methods; and 1,000 human judgments. The results of the SR are shown in Figure 6.

Bicubic as a traditional interpolation algorithm is undoubtedly the least effective, EDSR and WDSR are very similar, and SAN is much better, especially for the shape edges of the characters' SPGAN and SPSR yield a more realistic natural texture, and because SPSR introduces additional edge loss, the edges have higher definition and therefore higher perceived quality.

The denoising results are shown in Figure 7. FDNCNN, FFDNet, and IRCNN all remove the noise better, but the image information retention is poor, especially in terms of edge details. IPT is significantly better than FFDNet and IRCNN and has the highest PSNR. LIGN enhances the denoised image detail texture, which is lower than IPT on PSNR but has the best perceptual quality.

Tables 3 and 4 show the average scores for each restoration method in different IQAs. The results under RDIE have a similar ranking to MOS for both denoising and SR, while the other three differ, and in SR, the traditional metrics of the SPSR model are even worse than the interpolation method, which is clearly unreasonable. In denoising, one can still see similar ranking results for RDIE and MOS. Both SPSR and LIGN have a more natural content and structure, and although they may not be identical in detail to the reference image, they are clearly the best in terms of perceived quality, so they have the highest score under RDIE's evaluation.

Figure 8, this paper shows the subjective scores of 30 restored images and a scatter plot of the results under some IQA methods. The scattered points of our method are relatively clustered and concentrated, while the scattered points of others are very loose. SRCC and PLCC can prove that our method has a good rank correlation and linear correlation with MOS. The looseness of the scatter plot can be seen as the tightness of the relationship between the variables of the monotonic trend and the MOS. The tighter the scatter plot is, the stronger the relationship with the MOS is and the more consistent it is with the subjectively perceived quality of the image. Despite the widespread use of IQA methods such as PSNR, SSIM, and MS-SSIM, it is still difficult to achieve a high level of subjective and objective agreement when faced with image restoration methods that have special handling of detailed textures, especially GAN-based ones.

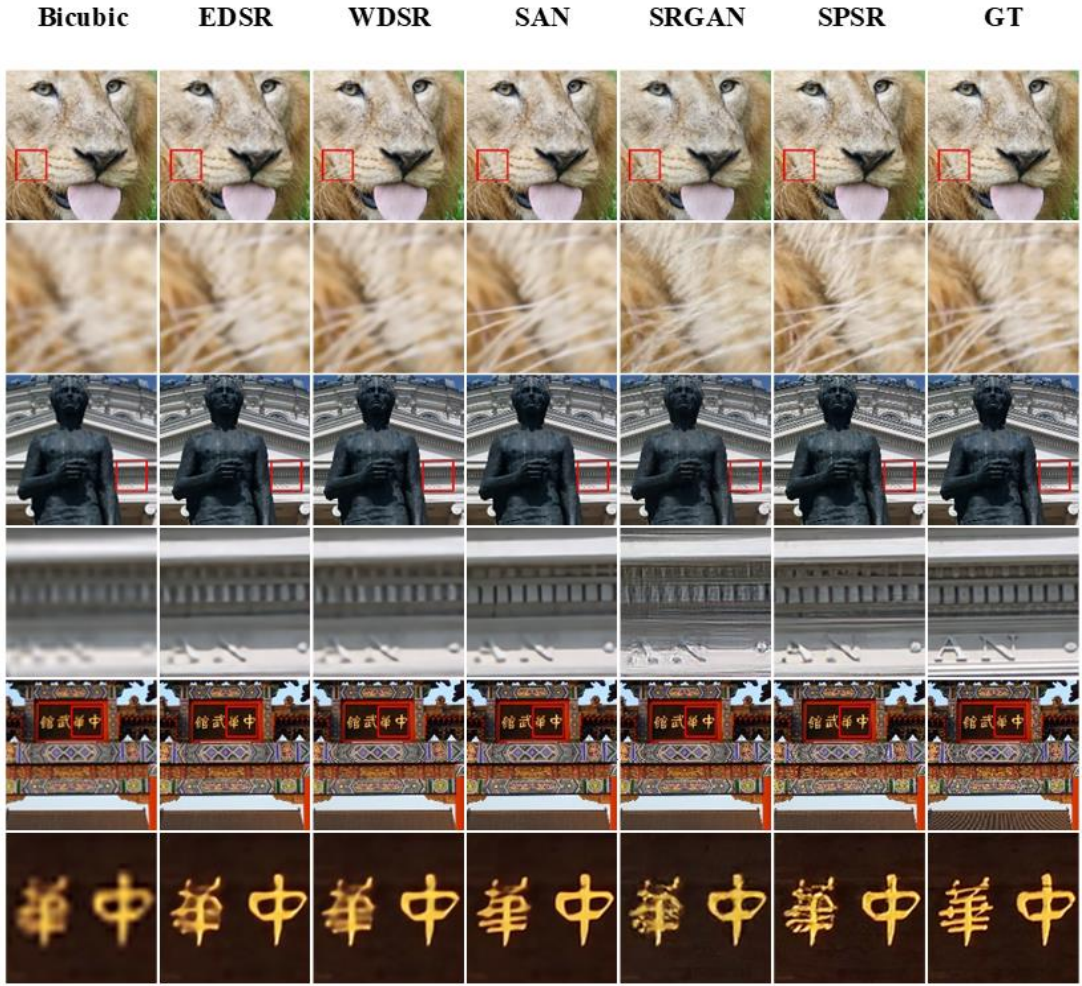

**Figure 6**. Results of different SR methods.

**Table 3**. Results of different IQA methods on our datasets (SR). The bold values are the best, and the superscripts indicate the ranking.

| Method | Bicubic | EDSR | WDSR | SAN | SRGAN | SPSR |
|---|---|---|---|---|---|---|
| PSNR↑ | 24.86 | 26.84 | 26.89 | **27.65** | 24.51 | 24.64 |
| SSIM↑[1] | 0.6962 | 0.7740 | 0.7745 | **0.7997** | 0.6773 | 0.6953 |
| MS-SSIM↑[9] | 0.8669 | 0.9171 | 0.9175 | **0.9297** | 0.8650 | 0.8769 |
| $RDIE_{5,32}$ ↓ | 41.53 | 33.73 | 32.89 | 32.59 | 26.33 | **23.14** |
| MOS↑ | 2.019 | 3.163 | 3.141 | 3.415 | 3.763 | **4.15** |

**Table 4**. Results of different IQA methods got our datasets (denoised). The bold values are the best, and the superscripts indicate the ranking.

| Method | FDNCNN | FFDNet | IRCNN | IPT | LIGN |
|---|---|---|---|---|---|
| PSNR↑ | 27.98 | 27.97 | 27.88 | **29.39** | 28.38 |
| SSIM↑[1] | 0.7916 | 0.7887 | 0.7898 | **0.8090** | 0.8066 |
| MS-SSIM↑[9] | 0.9334 | 0.9334 | 0.9312 | 0.9398 | **0.9406** |
| $RDIE_{5,32}$ ↓ | 25.96 | 28.41 | 25.57 | 25.92 | **23.93** |
| MOS↑ | 3.361 | 3.142 | 3.381 | 3.397 | **3.667** |

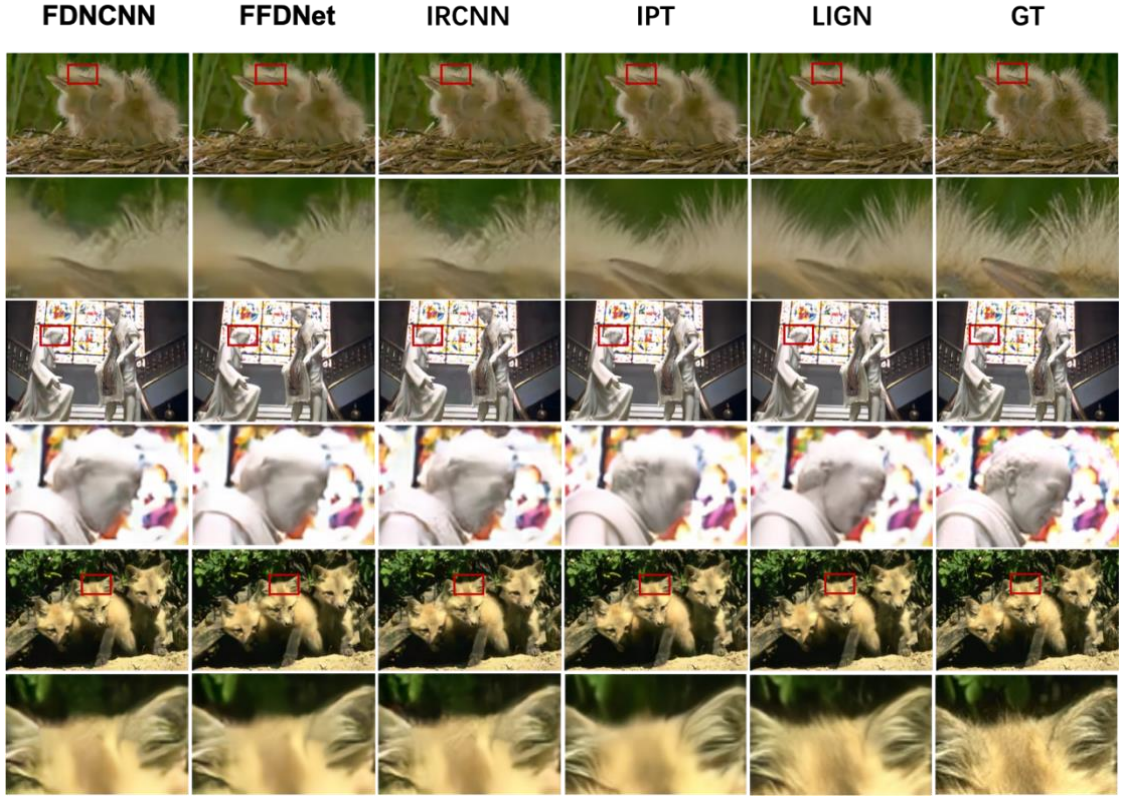

**Figure 7**. Results of different denoising methods.

*3.2. Results in PIPAL*

This paper uses the RDIE to compare with 15 other IQA methods. Other methods SRCC from benchmarks [37], results in Table 5. While the IFC performed well in its evaluation of traditional image restoration algorithms and PSNR-driven image restoration algorithms, it did not perform well against GAN-based image restoration algorithms, with poor correlation to subjective perceptual quality. With the above optimal parameters chosen, this paper has good performance for all three types of datasets, especially the GAN-based image restoration algorithm. Compared to other IQA methods, the RDIE can better measure the similar but not identical texture details generated based on GAN and has similar results to the perceived quality. In this study, three datasets were combined into a complete dataset of image restoration algorithms to calculate the SRCC. This method achieved the best performance for the dataset, with a significant improvement compared to the second-highest-performing SR-SIM.

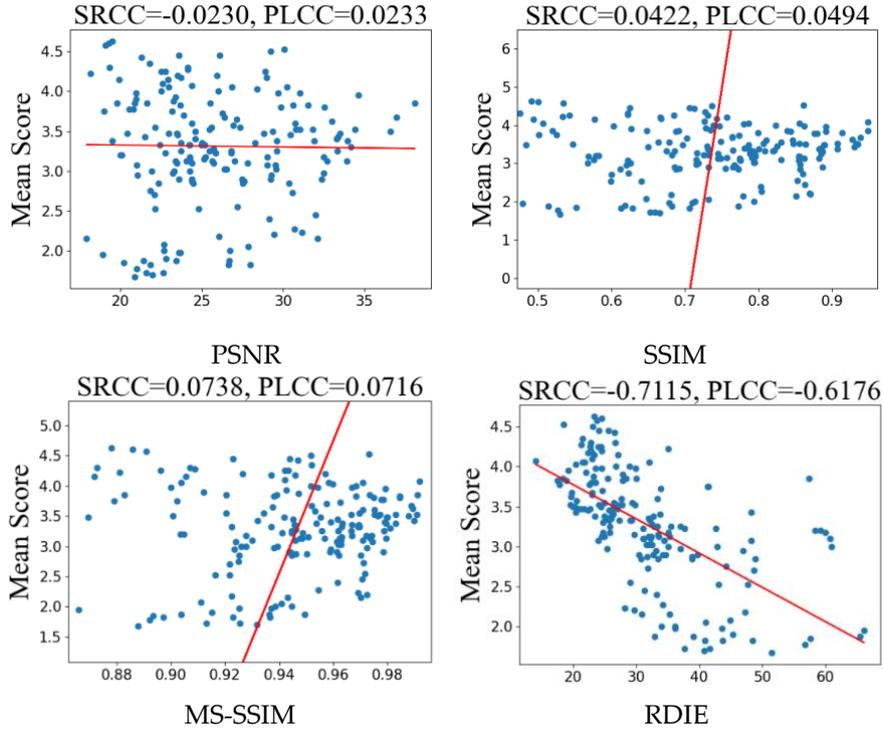

**Figure 8**. Analysis of different IQA methods.

**Table 5**. Results of different IQA algorithms for the PIPAL dataset. The upward arrows indicate that higher values are better for this metric and vice versa, with the best results marked in bold.

| Method | Traditional method | PSNR-oriented method | GAN-based method | All images Recovery method |
| --- | --- | --- | --- | --- |
| PSNR | 0.4782 | 0.5462 | 0.2839 | 0.4099 |
| NQM↑[38] | 0.5374 | 0.6462 | 0.3410 | 0.4742 |
| UQI↑[39] | 0.6087 | 0.7060 | 0.3385 | 0.5257 |
| SSIM↑[1] | 0.5856 | 0.6897 | 0.3388 | 0.5209 |
| MS-SSIM↑[9] | 0.6527 | 0.7528 | 0.3823 | 0.5596 |
| IFC↑[18] | **0.7062** | **0.8244** | 0.3217 | 0.5651 |
| VIF↑[19] | 0.6927 | 0.7864 | 0.3857 | 0.5917 |
| VSNR-FR↑[40] | 0.6146 | 0.7076 | 0.3128 | 0.5086 |
| RFSIM↑[41] | 0.4593 | 0.5525 | 0.2951 | 0.4232 |
| GSM↑[42] | 0.6074 | 0.6904 | 0.3523 | 0.5361 |
| SR-SIM↑[43] | 0.6561 | 0.7476 | 0.4631 | 0.6094 |
| FSIM↑[22] | 0.6515 | 0.7381 | 0.4090 | 0.5896 |
| $FSIM_c$ ↑[22] | 0.6509 | 0.7374 | 0.4058 | 0.5872 |
| VSI↑[44] | 0.6086 | 0.6938 | 0.3706 | 0.5475 |
| MAD↓[45] | 0.6720 | 0.7575 | 0.3494 | 0.5424 |
| RDIE↓ | 0.6476 | 0.7203 | **0.5280** | **0.6368** |

**4. Conclusions**

We propose a regional information entropy-based IQA method that is a reconstruction of the regional information entropy calculation process using a neural network approach. We validated the perceived ability of the RDIE method to restore images experimentally and determined the optimal RDIE parameter values for the image restoration task by ablating the window size, quantization level, and strides. We tested the proposed method using a dataset developed in this study and the PIPAL dataset. The results show that the proposed RDIE method is more responsive to subjective perceptions of images than other IQA measures, such as PSNR, SSIM, and MS-SSIM, and achieves better subjective and objective agreement.